\documentclass[journal]{IEEEtran}

\ifCLASSINFOpdf
\else
\fi

\usepackage{setspace}%????
\usepackage{bbm}
\usepackage[cmex10]{amsmath}
\usepackage{amssymb}
\usepackage{cite}
\usepackage{graphicx}
\usepackage{array,color}
\usepackage{amsmath}
\allowdisplaybreaks
\usepackage{stfloats}
\usepackage{graphicx}
\usepackage{epstopdf}
\usepackage{subfigure}
\usepackage{tabularx}
\usepackage{epsfig,epsf,color,balance,cite}
\usepackage{algorithmic}
\usepackage{algorithm}
\usepackage{url}
\usepackage{bm}
\usepackage{multirow}

\usepackage{amsthm}

\ifodd 1
\usepackage{soul}
\usepackage{color}
\setstcolor{red}

\newcommand{\del}[1]{\st{#1}} %deleting the text

\newcommand{\com}[1]{\textbf{\color{red} (COMMENT: #1)}} %comment of the text
\newcommand{\response}[1]{\textbf{\color{green} (RESPONSE: #1)}} %response to comment
\else

\newcommand{\del}[1]{}

\newcommand{\com}[1]{}
\newcommand{\comg}[1]{}
\newcommand{\response}[1]{}
\fi
\title{\huge {Efficient Channel Estimation for Rotatable Antenna-Enabled Wireless Communication}}
\author{ 
Xue Xiong, Beixiong Zheng,~\IEEEmembership{Senior Member,~IEEE},  Wen Wu,~\IEEEmembership{Senior Member,~IEEE}, Xiaodan Shao, \IEEEmembership{Member,~IEEE}, Liang Dai, Ming-Min Zhao,~\IEEEmembership{Senior Member,~IEEE},
and Jie Tang,~\IEEEmembership{Senior Member,~IEEE} 
\vspace{-1cm}
\thanks{			
X. Xiong is with the School of Future Technology, South China University of Technology, Guangzhou 511442, China, and also with the Frontier Research Center, Peng Cheng Laboratory, Shenzhen 518055, China (e-mail: ftxuexiong@mail.scut.edu.cn).

B. Zheng and L. Dai are with the School of Microelectronics, South China University of Technology, Guangzhou 511442, China (e-mail: bxzheng@scut.edu.cn; 202321061996@mail.scut.edu.cn).
			
W. Wu is with the Frontier Research Center, Peng Cheng Laboratory, Shenzhen 518055, China (e-mail: wuw02@pcl.ac.cn).

X. Shao is with the Department of Electrical and Computer Engineering, University of Waterloo, Waterloo, ON N2L 3G1, Canada (e-mail: x6shao@uwaterloo.ca).

M. M. Zhao is with the College of Information Science and Electronic Engineering, %and also with Zhejiang Provincial Key Laboratory of Information Processing, Communication and Networking (IPCAN),
Zhejiang University, Hangzhou 310027, China (e-mail: zmmblack@zju.edu.cn).
			
J. Tang is with the School of Electronic and Information Engineering, South China University of Technology, Guangzhou 510640, China (e-mail: eejtang@scut.edu.cn).

		}
}

\begin{document}
\maketitle
\begin{abstract}	
Rotatable antenna (RA) is a promising antenna architecture that exploits additional spatial degrees of freedom (DoFs) to enhance the communication performance. 
To fully obtain the performance gain provided by RAs, accurate channel state information (CSI) is essential for adjusting the orientation/boresight of each antenna.
In this letter, we propose an efficient channel estimation scheme for RA communication systems, where the base station (BS) can sequentially and adaptively adjust the orientations of RAs to enrich the environmental observations from diverse angular perspectives, thereby enhancing the channel estimation accuracy.
The proposed scheme includes two main procedures that are conducted alternately during each channel training period. 
Specifically, the first procedure is to estimate the CSI with given RAs' orientations, involving the angle-of-arrivals (AoAs) information and path gains.
Then, based on the estimated CSI, the second procedure adjusts the RAs' orientations to maximize the effective channel gain.  
Simulation results demonstrate that the proposed channel estimation method outperforms other benchmark schemes.
\end{abstract}
\IEEEpeerreviewmaketitle

\vspace{-0.4cm}
\section{Introduction} 
To fulfill the increasingly stringent requirements of future sixth-generation (6G) wireless networks, such as sub-millisecond access latency and terabit-per-second (Tbps) data rate, multiple-input multiple-output (MIMO) stands out as one of the key technologies to significantly enhance the capacity and reliability of wireless communications \cite{Shen2022Holistic,Extremely2024Wang}.
However, traditional MIMO and/or massive MIMO systems mainly rely on fixed antenna-based configurations, which cannot fully utilize the spatial degree of freedoms (DoFs) to
achieve better communication and sensing performance.
To overcome this problem, movable antenna (MA)/fluid antenna system (FAS) has emerging as a promising solution for enhancing communication performance by dynamically adjusting antenna positions within predefined spatial regions to achieve better channel conditions \cite{Zhu2024Modeling,New2024Tutorial}.
Moreover, six-dimensional movable antenna (6DMA) systems are endowed with more spatial DoFs to enhance network performance by enabling adjustments of three-dimensional (3D) positions and 3D rotations of antenna arrays \cite{Shao20256D}.
However, despite their enhanced adaptability to dynamic wireless environments, the increased implementation complexity of 6DMA systems poses significant challenges due to the extensive modifications required for existing base station (BS) infrastructures. 
 \begin{figure}[!t]
 	\centering		
 	\includegraphics[width=3.0in]{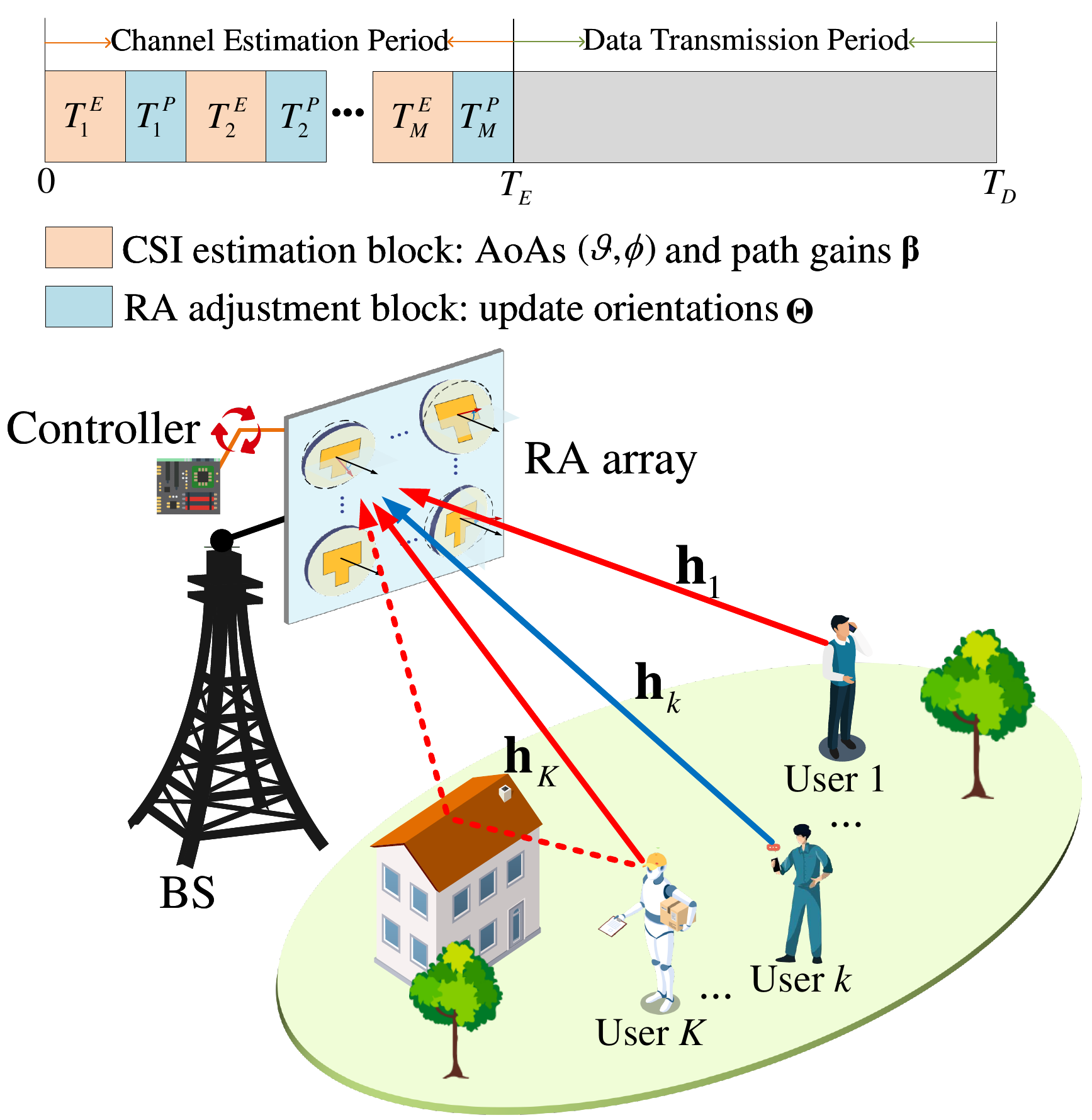}
 	\caption{RA-enabled multi-user uplink communication system.}
 	\vspace{-0.7cm}
 	\label{system}
 \end{figure}

Recent advances in rotatable antenna (RA) systems enable dynamic boresight adjustment of individual antennas, allowing for customizable directional gain patterns with compact hardware configurations \cite{Beixiong2025rotatableantenna,wu2024modeling,zheng2025rotatable}. 
This architecture provides additional DoFs, thereby facilitating several studies that have demonstrated the great potential of RAs in diverse wireless systems \cite{wu2024modeling,zheng2025rotatable,zhou2025rotatable,dai2025rotatable}. %\cite{wu2024modeling,zheng2025rotatable,zhou2025rotatable,xie2025thz,dai2025rotatable}.
Foundational studies \cite{wu2024modeling,zheng2025rotatable} established RA system models and quantified performance gains, while subsequent works extended RA capabilities to integrated sensing and communication (ISAC) \cite{zhou2025rotatable} and physical-layer security \cite{dai2025rotatable}, demonstrating significant improvements in weighted performance metrics and secrecy rates.
Prior works on RA primarily concentrate on the analysis and optimization of their communication performance, which highly relies on the acquisition of accurate channel state information (CSI).
This thus calls for efficient CSI estimation methods tailored to the RA-enabled systems. In contrast to traditional fixed antenna systems, RAs present distinctive advantages to enhance channel estimation by actively exploiting the additional spatial DoFs introduced by antenna boresight rotation.
Specifically, by sequentially or adaptively scanning the propagation environment, RAs can capture multi-perspective channel measurements, thereby enabling more accurate multipath resolutions.
Consequently, the active reception mechanism inherent in RA-based systems opens up a new road to improve the accuracy of channel estimation or reduce the pilot overhead.

Motivated by the above, we propose an efficient channel estimation scheme for RA-enabled wireless communication system. The proposed scheme consists of two main procedures that are conducted alternately during each channel training period. 
The first procedure involves estimating the CSI, including angle-of-arrivals (AoAs) information and path gains.
This is accomplished by employing a multiple signal classification (MUSIC) algorithm for AoAs estimation and the least squares (LS) method for path gains estimation. In the second procedure, the estimated CSI is utilized to help adjust the RAs' orientations, thereby maximizing the effective channel gain between the BS and users. Simulation results demonstrate that the proposed channel estimation method achieves a superior performance gain compared to other benchmark schemes.

\vspace{-0.2cm}
\section{System Model}\label{sys}
%\begin{figure}[!t]
%	\centering		
%	\includegraphics[width=3.0in]{fig_system4.pdf}
%	\caption{RA-enabled multi-user uplink communication system.}
%	\vspace{-0.5cm}
%	\label{system}
%\end{figure}
As shown in Fig. \ref{system}, we consider an RA-enabled uplink communication system, where $K$ users simultaneously communicate with the BS.
The BS is equipped with a uniform planar array (UPA) consisting of $N$ directional RAs, while each user is equipped with only a single isotropic antenna.
%are distributed in the AOI. Each user is equipped with a single isotropic fixed-position antenna (FPA) and simultaneously transmits its signals to the BS equipped with a uniform planar array (UPA) consisting of $N$ directional RAs. 
Without loss of generality, we assume that the UPA is placed on the $x$-$y$ plane of a 3D Cartesian coordinate system with $N\triangleq N_xN_y$, where $N_x$ and $N_y$ denote the numbers of RAs along $x$- and $y$-axes, respectively.
For ease of exposition, the reference position of the UPA is assumed to be at the origin.
Denote the distance between the BS and the $k$-th user as $r_k$, and represent the position of the $k$-th user as $\mathbf{q}_k=\left[r_k\Phi_k, r_k\Psi_k,r_k\Omega_k \right]^T\in \mathbb{R} ^{3\times 1}$ with $\Phi_k=\text{sin}\vartheta_k \text{sin}\phi_k$, $\Psi_k=\text{cos}\vartheta_k$, and $\Omega_k=\text{sin}\vartheta_k \text{cos}\phi_k$, where $\vartheta_k\in \left[ -\frac{\pi}{2},\frac{\pi}{2}\right]$ and $\phi_k\in [0,\pi]$ denote the elevation angle and azimuth angle of the $k$-th user with respect to the origin of the coordinate system, respectively, for $k=1,...,K$.  

The original orientations/boresights of all RAs are assumed to be parallel to the $z$-axis, and the boresight of each RA can be independently adjusted in 3D direction mechanically and/or electrically through a common smart controller \cite{zheng2025rotatable}. 
Specifically, the 3D orientation adjustment of each RA can be described by two deflection angles, i.e., the zenith angle $\theta_n^{\mathrm{z}}$ and the azimuth angle $\theta_n^{\mathrm{a}}$.
To characterize the 3D orientation 
of each RA, a pointing vector associated with the zenith angle $\theta_n^{\mathrm{z}}$ and azimuth angle $\theta_n^{\mathrm{a}}$ of the $n$-th RA is defined as 
\begin{equation}
	\mathbf{f}(\boldsymbol{\theta}_n)=\left[\text{sin}(\theta_n^{\mathrm{z}})\text{cos}(\theta_n^{\mathrm{a}}),\text{sin}(\theta_n^{\mathrm{z}})\text{sin}(\theta_n^{\mathrm{a}}), \text{cos}(\theta_n^{\mathrm{z}}) \right]^T,
\end{equation}
where $\boldsymbol{\theta}_n=[\theta_n^{\mathrm{z}},\theta_n^{\mathrm{a}}]^T$ denotes the deflection angle vector of the $n$-th RA, and we have $\left\| \mathbf{f}(\boldsymbol{\theta}_n) \right\| =1 $ for normalization.
To account for the antenna boresight adjustment range, the zenith angle of each RA should be constrained within a specific range:
% i.e., $ 0	\le \theta_{n}^{\mathrm{z}} \le \theta_{\text{max}}, \forall n$, 
\begin{equation}
   0	\le \theta_{n}^{\mathrm{z}} \le \theta_{\text{max}}, \forall n,
\end{equation}
where $\theta_{\text{max}} \in \left[0, \frac{\pi}{2} \right] $ represents the maximum zenith angle that each RA is allowed to adjust. 
\subsection{Channel Model}
\vspace{-0.2cm}
We assume that all the links between the BS and users are light-of-sight (LoS)\footnote{In this paper, the core concept of the proposed channel estimation scheme in Section III is to estimate the critical parameters of the channel, such as the AoA and path gain for each resolvable channel path. Therefore, this method can be extended to more general environment with scatters and multiple propagation paths by treating each path independently and estimating its AoA and path gain.}.
For convenience, we first define a one-dimensional (1D) steering vector function for a generic uniform linear
array (ULA) as follow:
\begin{equation}
	\boldsymbol{e}(\psi,\bar{N})\triangleq \left[ 1,e^{j\frac{2\pi d}{\lambda} \psi},...,e^{j \frac{2\pi d}{\lambda} (\bar{N}-1) \psi}\right]^T \in \mathbb{C}^{\bar{N} \times 1},
\end{equation}
where $j=\sqrt{-1}$ denotes the imaginary unit, $d$ denotes the antenna spacing, $\lambda$ is the signal wavelength, $\psi$ denotes the constant phase-shift difference between the signals at two adjacent antennas, and $\bar{N}$ denotes the number of antennas in the ULA. 
In practice, since the propagation distances between the BS and users are much greater than the physical size of the UPA, the propagation channels can be well characterized by the far-field model with parallel wavefronts. 
Under the UPA model, we let $\mathbf{a}_k(\vartheta_k,\phi_k)\in \mathbb{C}^{N\times1}$ denote the array response vector associated with 
AoA/angle-of-departure (AoD) pair $(\vartheta_k,\phi_k)$ of the $k$-th user, and it can be expressed as 
\begin{equation}
	\begin{aligned}
		\mathbf{a}_k(\vartheta_k,\phi_k)= &\boldsymbol{e}\left( \text{cos}(\phi_k)\text{cos}(\vartheta_k),N_x  \right) 
		\\	&
		\otimes \boldsymbol{e}\left( \text{cos}(\phi_k)\text{sin}(\vartheta_k),N_y \right),
	\end{aligned}
\end{equation}
where $\otimes$ denotes the Kronecker product.

Note that the effective antenna gain for each RA highly depends on both signal arrival/departure angle and antenna gain pattern, which characterizes the antenna radiation power distribution over different directions. 
Let $G(\epsilon,\phi)$ denote the antenna gain pattern, which is described by a generic cosine pattern model as follows \cite{balanis2016antenna}.
%For the antenna gain pattern, we consider a generic cosine pattern model as follow:
\begin{equation}
G(\epsilon,\phi)=\begin{cases}
	G_0 \text{cos}^{2p}(\epsilon),\epsilon \in [0,\pi/2 ), \phi \in [0,2\pi ) \\
	0,  \text{otherwise,}
\end{cases}
\end{equation}
where $\epsilon$ and $\phi$ are the zenith and azimuth angles of any spatial directions with respect to the RA's boresight direction, $p$ determines the directivity of antenna, and $G_0$ is the maximum gain in the boresight direction (i.e., $\epsilon=0$) with $G_0=2(2p+1)$ satisfying the law of power conservation. 
Thus, the directional gain between the $n$-th RA and the $k$-th user is represented by $G_{k,n}(\boldsymbol{\theta}_n)={G_0 \text{cos}^{2p}(\epsilon_{k,n})}$, where $\text{cos}(\epsilon_{k,n})=\mathbf{f}^T(\boldsymbol{\theta}_n) \vec{\mathbf{q}}_{k}$ is the projection between the $k$-th user direction vector %$\vec{\mathbf{q}}_{k} \triangleq \frac{\mathbf{q}_k-\mathbf{w}}{\left\|\mathbf{q}_k-\mathbf{w} \right\|}$
$\vec{\mathbf{q}}_{k} \triangleq \frac{\mathbf{q}_k}{\left\|\mathbf{q}_k\right\|}=
[{\Phi _k},{\Psi _k},{\Omega _k}]^T$
and the point vector of the $n$-th RA.
Here, $\lVert\cdot\rVert$ denotes the Euclidean norm of a vector.
Following that, we can determine the RA radiation coefficients for all elements and encapsulate them in a vector $\mathbf{g}_k(\boldsymbol{\Theta})\in\mathbb{C}^{N\times 1}$, which is given by 
\begin{equation} \label{radiation coefficient}
	\mathbf{g}_k(\boldsymbol{\Theta})=\left[ \sqrt{G_{k,1}(\boldsymbol{\theta}_1)},...,\sqrt{G_{k,N}(\boldsymbol{\theta}_N)}\right] ^T,
\end{equation}
where  $\mathbf{\Theta}=[\boldsymbol{\theta}_1,...,\boldsymbol{\theta}_N]\in \mathbb{R}^{2\times N}$ is the orientation matrix of RAs.
Therefore, the channel model between the BS and the $k$-th user can be expressed as 
\begin{equation}\label{channel model}
	\begin{aligned}
	\mathbf{h}_k(\mathbf{\Theta}) =  \beta_k \mathbf{g}_{k}(\mathbf{\Theta}) \odot  \mathbf{a}_k(\vartheta_k,\phi_k),\\
	\end{aligned}
\end{equation}
where $\beta_k= \frac{\lambda}{4\pi r_k}e^{-j\frac{2\pi}{\lambda}r_k}$ denotes the complex-valued channel path gain between the BS and the $k$-th user, $\odot$ represents the Hadamard product.

\vspace{-0.3cm}
\subsection{Signal Model}
Let $\mathbf{s}(t)=[s_1(t),...,s_K(t)]^T \in \mathbb{C}^{K\times 1}$ be the transmit pilot signal of all users at time instant $t$ and further assume that $\mathbb{E}[|{s}_k(t)|^2]=p_k$, where $p_k $ is the transmit power of the $k$-th user. 
Then, the received signal at BS can be expressed as 
\begin{equation}\label{received_signal}
	\begin{aligned}
		\mathbf{y}^{[t]}(\mathbf{\Theta}) &=\sum_{k=1}^{K}  \mathbf{h}_k(\mathbf{\Theta}) s_k(t) + \mathbf{n}(t)\\
		&=\mathbf{G}(\mathbf{\Theta})\odot\mathbf{A}(\boldsymbol{\vartheta},\boldsymbol{\phi})\text{diag}(\boldsymbol{\beta})
		\mathbf{s}(t)+\mathbf{n}(t),
	\end{aligned}
\end{equation} 
where $\mathbf{G}(\mathbf{\Theta})=\left[\mathbf{g}_1(\mathbf{\Theta}),...,\mathbf{g}_K(\mathbf{\Theta})\right] \in \mathbb{R}^{N\times K}$ denotes the directional gain matrix, and $	\mathbf{A}(\boldsymbol{\vartheta},\boldsymbol{\phi})=\left[ \mathbf{a}_1(\vartheta_1,\phi_1),...,\mathbf{a}_K(\vartheta_K,\phi_K)\right] 
\in \mathbb{C}^{N\times K}$
%\begin{equation}\label{RA_channel}
%	\mathbf{A}(\boldsymbol{\vartheta},\boldsymbol{\phi})=\left[ \mathbf{a}_1(\vartheta_1,\phi_1),...,\mathbf{a}_K(\vartheta_K,\phi_K)\right] 
%	\in \mathbb{C}^{N\times K}
%\end{equation} 
is the array manifold matrix that collects all the array response vectors corresponding to $K$ users, $\boldsymbol{\vartheta}=[\vartheta_1,...,\vartheta_K]^T\in \mathbb{R}^{K\times 1}$, $\boldsymbol{\phi}=[\phi_1,...,\phi_K]^T\in \mathbb{R}^{K\times 1}$,  $\boldsymbol{\beta}=[\beta_1,...,\beta_K]^T\in \mathbb{C}^{K\times 1}$,
and $\mathbf{n}(t)$ denotes the zero-mean additive white Gaussian noise (AWGN) with variance $\sigma^2$. 
Furthermore, we denote the sum channel gain between the BS and users as $\gamma$, i.e.,
\begin{equation}\label{SNR}
	\begin{aligned}
		\gamma=\sum_{k=1}^{K} \left\| \mathbf{h}_k (\mathbf{\Theta}) \right\| ^2. 
	\end{aligned}
\end{equation}
Notably, the overall channel gain is influenced by the channel condition, which is characterized by the orientations of RAs $\mathbf{\Theta}$, as well as the path gains $\boldsymbol{\beta}$ and AoAs information $ (\boldsymbol{\vartheta},\boldsymbol{\phi}) $ that are only determined by the environmental factors.

\section{Proposed Channel Estimation Scheme}
%Generally, one transmission frame includes two sub-frames: channel estimation and data transmission.
In this section, we propose an efficient channel estimation scheme to facilitate subsequent data transmission in the RA-enabled wireless communication system, as illustrated in Fig.~\ref{system}. 
The proposed estimation scheme consists of two main procedures that are conducted alternatively during each channel training period.
Specifically, the duration $T_E$ of each channel training period is divided into $M$ blocks, with each block further partitioned into two sub-blocks: CSI estimation sub-block $T_{m}^{E}$ and RAs' orientations adjustment sub-block $T_{m}^{P}$, with $m=1,...,M$.
First, in the CSI estimation sub-block with preset orientations of RAs\footnote{During the initial channel estimation without any reference CSI, the BS can preset orientations of RAs, such as utilizing a conventional fixed antenna to received pilots, i.e., $\mathbf{\Theta}=\mathbf{0}$. For the subsequent (non-initial) estimation procedure, the BS can utilize the latest updated CSI to set its RAs' orientations to maximize the overall channel gain for enhancing the channel estimation performance.}, the BS employs the MUSIC algorithm to determine the AoA information $(\boldsymbol{\vartheta},\boldsymbol{\phi})$ and utilizes the LS method to estimate the path gain $\boldsymbol{\beta}$. 
Then, in the RAs' orientations adjustment sub-block, the BS leverages the estimated CSI obtained in previous procedure to adjust the RAs' orientations design, thereby enhancing the sum channel gain. 
In the following, we elaborate the above two procedures in detail, respectively.
%\vspace{-0.6cm}
\subsection{CSI Estimation}
\textbf{1) AoAs Estimation}: Based on the received signal in \eqref{received_signal} associated with the orientation matrix, existing spectral-based algorithms such as  MUSIC or estimation of signal parameters via rotational invariance technique (ESPRIT) \cite{Schmidt1986Multiple} can be applied to estimate the AoAs  $(\boldsymbol{\vartheta},\boldsymbol{\phi})$.
By exploiting $T_{m}^{E}$ time slots in the $m$-th CSI estimation sub-block, the array covariance matrix of the received signal can be calculated as 
\begin{equation}
	\mathbf{R}(\mathbf{\Theta})=\mathbb{E}\left\lbrace
	\mathbf{y}^{[t]}(\mathbf{\Theta})(\mathbf{y}^{[t]}(\mathbf{\Theta}))^H \right\rbrace =\frac{1}{T_m^E}\sum_{t=1}^{T_m^E}  \mathbf{y}^{[t]}(\mathbf{\Theta})(\mathbf{y}^{[t]}(\mathbf{\Theta}))^H.
\end{equation}
For convenience, assume that the value of $K$ is known or correctly estimated by the Akaike information criterion (AIC) or the minimum description length (MDL) detection criterion \cite{zhang1989statistical}.
After conducting the eigenvalue decomposition, we can obtain
\begin{equation}
	\mathbf{R}(\mathbf{\Theta})=\mathbf{E}_s \mathbf{\Sigma}_s\mathbf{E}_s^H+\mathbf{E}_n \mathbf{\Sigma}_n\mathbf{E}_n^H,
\end{equation}
where $\mathbf{\Sigma}_s\in \mathbb{C}^{K\times K}$ and $\mathbf{\Sigma}_n\in \mathbb{C}^{(N-K)\times (N-K)}$ are diagonal matrices having the $K$ largest and $(N-K)$ smallest eigenvalues of $\mathbf{R}$ on their diagonal, respectively. 
The matrices $\mathbf{E}_s\in \mathbb{C}^{N\times K}$ and $\mathbf{E}_n\in \mathbb{C}^{N\times (N-K)}$ contains the eigenvectors corresponding to eigenvalues on the diagonal of $\mathbf{\Sigma}_s$ and $\mathbf{\Sigma}_n$, respectively. 
By utilizing the orthogonal relationship between the signal subspace and the noise subspace, the spectrum function of signal sources can be presented by 
\begin{equation}\label{spectrum} 
	V(\vartheta,\phi;\mathbf{\Theta}) =\frac{1}{ \mathbf{b}^H(\vartheta,\phi;\mathbf{\Theta})\mathbf{E}_n \mathbf{E}_n^H \mathbf{b}(\vartheta,\phi;\mathbf{\Theta})},
\end{equation}
where $\mathbf{b}(\vartheta,\phi;\mathbf{\Theta})=\mathbf{g}(\mathbf{\Theta})\odot\mathbf{a}(\vartheta,\phi)$. 
Taking the directional gain pattern into consideration, the maximum spatial spectrum peaks are not only relevant to the potential user direction $(\vartheta,\phi)$, but also associated with the orientations of RAs $\mathbf{\Theta}$ commonly shared by $K$ users. 
The AoAs of all $K$ users are obtained by finding top-$K$ peaks in the pseudo-spectrum (\ref{spectrum}).

\textbf{2) Path Gains Estimation}: For the considered RA-enabled wireless communication system, the orientation matrix ${\mathbf{\Theta}(t)}$ are generally set as a constant in each CSI estimation sub-block, i.e., $\mathbf{\Theta}(t)=\mathbf{\Theta}, t=1,...,T_m^E$. Consequently, the received signal model for $T_m^E$ time slots is given by
\begin{equation}\label{received signal}
	\mathbf{Y}=\mathbf{B}(\mathbf{\Theta})\mathrm{diag}(\boldsymbol{\beta})\mathbf{S}+\mathbf{N} \in \mathbb{C}^{N\times T_m^E},
\end{equation}
where $\mathbf{B}(\mathbf{\Theta})= \mathbf{G}(\mathbf{\Theta}) \odot {\mathbf{A}}({\boldsymbol{\vartheta}},{\boldsymbol{\phi}}) \in \mathbb{C}^{N\times K}$, %$\boldsymbol{\beta} \in \mathbb{C}^{K\times 1}$, 
$\mathbf{S}=\left[ \mathbf{s}(1),...,\mathbf{s}(T_m^E)\right] \in \mathbb{C}^{K\times T_m^E}$, $\mathbf{N}=\left[\mathbf{n}(1),...,\mathbf{n}(T_m^E)\right]  \in \mathbb{C}^{N\times T_m^E}$ denotes the AWGN matrix.
To facilitate further analysis, we reshape the received signal matrix $\mathbf{Y}$ into a $NT_m^E \times 1$ dimensional vector $\tilde{\mathbf{y}}$:
\begin{equation}
	\tilde{\mathbf{y}}=\mathbf{X} \boldsymbol{\beta} +\tilde{\mathbf{n}}= \underbrace{\begin{bmatrix}
			{\mathbf{B}}\text{diag}\left( \mathbf{s}(1)\right)   \\
			{\mathbf{B}}\text{diag}\left( \mathbf{s}(2)\right) \\
			\vdots \\
			{\mathbf{B}}\text{diag} \left( \mathbf{s}(T_m^E)\right) \\
	\end{bmatrix}}_{\mathbf{X}\in \mathbb{C}^{NT_m^E\times K}}\boldsymbol{\beta}+\tilde{\mathbf{n}}.
\end{equation}
Then, based on the pre-designed orientations of RAs in previous procedures, the LS method can be applied to estimate the path gains as follows:
\vspace{-0.1cm}
\begin{equation}\label{path gain estimation}
	\hat{\boldsymbol{\beta}}= (\mathbf{X}^H\mathbf{X})^{-1}\mathbf{X}^H \tilde{\mathbf{y}}.
\end{equation}

\vspace{-0.65cm}
\subsection{RA Orientation Adjustment}
During the RAs' orientations adjustment sub-block $T_m^P$, based on the estimated CSI, i.e., ($ \hat{\boldsymbol{\vartheta}}$, $\hat{\boldsymbol{\phi}}$, $\hat{\boldsymbol{\beta}}$) in the sub-block $T_m^E$, the orientations of RAs are optimized at the BS to maximize the overall channel gain as defined in \eqref{SNR}. 
The associated optimization problem can be formulated as
\begin{subequations}\label{average_optimize}
	\begin{align}
		&\underset{\bf{\Theta}}{\max} \quad \gamma
		=\sum_{k=1}^{K} \left\| \mathbf{h}_k (\mathbf{\Theta}) \right\| ^2\\
		&{\rm{s.t.}} \quad 0\le {\theta_{n}^{\mathrm{z}}} \le {\theta _{\max }},\forall n.
	\end{align}
\end{subequations}

Note that the point vector $\mathbf{f}(\boldsymbol{\theta}_n)$ is essentially a unit vector on the unit sphere, and the RA deflection angles mainly affect the effect channel power gains through the projections $\text{cos}(\epsilon_{k,n})=\mathbf{f}^T(\boldsymbol{\theta}_n) \vec{\mathbf{q}}_{k}$.
For ease of subsequent derivation, we introduce an auxiliary variable $\mathbf{f}_n$ to equivalently replace the influence of deflection angle vector $\boldsymbol{\theta}_{n}$ on the point vector of $n$-th RA, i.e., $\mathbf{f}_n=\mathbf{f}(\boldsymbol{\theta}_{n})$.
Accordingly, the overall channel gain can be reformulated as
\begin{equation}
	\gamma=\sum_{k=1}^{K}\sum_{n = 1}^{N}  \left| \beta_k \right|^2 {G_0\text{cos}^{2p}(\epsilon_n)}
	=\sum_{k = 1}^{K}\sum_{n = 1}^{N} \mu_k({\bf{f}}^T_n{\vec{\mathbf{q}}}_{k})^{2p},
\end{equation}
where $ \mu_k= {\left| \beta_k \right|^2 G_0} $.
Let $\mathbf{F}\triangleq [\mathbf{f}_1,\mathbf{f}_2,...,\mathbf{f}_N]$, problem \eqref{average_optimize} can be reformulated as
\begin{subequations}\label{Obejective function1}
	\begin{align}
		\underset{\bf{F}}{\max} &\quad \mathcal{L}({\mathbf{F}})= \sum_{n = 1}^N \sum_{k = 1}^K \mu_k ({\bf{f}}^T_n{\vec{\mathbf{q}}}_{k})^{2p}\\
		{\rm{s.t.}} &\quad {\bf{f}}_n^T{{\mathbf{e}}} \ge \cos ({\theta_{{\rm{max}}}}), \forall n,\\
		& \quad \left\| {\bf{f}}_n \right\| = 1,\forall n,
	\end{align}
\end{subequations}
where $\mathbf{e}=[0,0,1]^T$. The constraint in (18b) is equivalent to (16b), and (18c) ensures that $\mathbf{f}_n$ is a unit vector.

Exploiting the objective function in (18a), problem \eqref{Obejective function1} can be decomposed into $N$ subproblems, each of which independently optimizes the deflection angle vector of one RA. 
For the $n$-th RA, the corresponding subproblem is given by
\begin{subequations}\label{Obejective function2}
	\begin{align}
		\underset{\mathbf{f}_n}{\max} &\quad  \tilde{\mathcal{L}}({\mathbf{f}_n})= \sum_{k = 1}^K  \mu_k ({\bf{f}}^T_n{\vec{\mathbf{q}}}_{k})^{2p}\\
		{\rm{s.t.}} &\quad {\bf{f}}_n^T{{\bf{e}}} \ge \cos ({\theta_{{\rm{max}}}}), \\
		& \quad \left\| {\bf{f}}_n \right\| = 1,\forall n.
	\end{align}
\end{subequations}
Since problem \eqref{Obejective function2} is still non-convex, we utilize the iterative gradient ascent approach to solve it. 
First, we compute the gradient of $\tilde{\mathcal{L}} (\mathbf{f}_n)$ as 
\begin{equation}
	\nabla_{{\mathbf{f}}_n} \tilde{\mathcal{L}} (\mathbf{f}_n)
	=2p\sum_{k = 1}^K \mu_k {({\bf{f}}^T_n{\vec{\bf{q}}_{k}})^{2p - 1}}{\vec{\bf{q}}}^T_{k}.
\end{equation}
Then, in the $i$-th iteration, the point vector is updated as
\begin{equation}
	{\mathbf{f}}_n^{(i)}= \Pi_{\mathcal{C}} \left({\mathbf{f}}_n^{(i-1)}+\xi \nabla_{{\mathbf{f}}_n} \tilde{\mathcal{L}} ({\mathbf{f}}_n^{(i-1)})\mid _{{\mathbf{f}}_n={\mathbf{f}}_n^{(i-1)}}\right),
\end{equation}
where $\xi$ denotes the step length, and $\Pi_{\mathcal{C}}$ denotes the projection onto the feasible region defined by constraints (19b) and (19c).
Specifically, if the point vector $\mathbf{f}_n$ violates the zenith angle constraint, i.e.,  $\mathbf{f}_n^T\mathbf{e} < \cos(\theta_{\rm{max}})$, then $\hat{\mathbf{f}}_n = \mathbf{f}_n-(\mathbf{f}_n^T\mathbf{e}- \cos ({\theta_{{\rm{max}}}}))\mathbf{e}$. Subsequently, normalization is performed by setting $\tilde{\mathbf{f}}_n = \frac{\hat{\mathbf{f}}_n}{\|\hat{\mathbf{f}}_n\|}$.

\vspace{-0.4cm}
\subsection{Computational Complexity Analysis}
For the $m$-th block of each channel training period, the computational complexity of the proposed channel estimation method consists of three main components: AoAs estimation $\mathcal{O}_{AoA}$, path gains estimation $\mathcal{O}_{gain}$, and RAs' orientations adjustment $\mathcal{O}_{RA}$.
Specifically, the complexity of the MUSIC algorithm is 
$\mathcal{O}_{AoA}=\mathcal{O}(N^2T_m^E+N^3+LN(N-K))$, where $L$ is the number of search directions.
The path gain estimation in \eqref{path gain estimation} has a computational complexity of $ \mathcal{O}_{gain}= \mathcal{O}(NK^2T_m^E+NKT_m^E+K^3+K^2)$.
For the RAs' orientations adjustment, the computational complexity is $\mathcal{O}_{RA}=\mathcal{O}(3N/\epsilon^2)$ for a given solution accuracy of $\epsilon$. 
Accordingly, the total computational complexity is $M(\mathcal{O}_{AoA}+\mathcal{O}_{gain}+\mathcal{O}_{RA})$.

\vspace{-0.cm}
\section{Simulation Results}
In this section, we provide the simulation results to evaluate the performance of the proposed channel estimation scheme for the RA-enabled wireless communication system.
For ease of exposition, we assume that the BS 
%equipped with a RA-based ULA
and the users are located on the same two-dimensional (2D) plane, and thus we only need to focus on the AoAs $\left\lbrace \vartheta_k \right\rbrace$.
In the simulations, $K=3$ users are considered, with their respective AoAs set as
$\vartheta_1=15.4^\circ$, $\vartheta_2=30.7^\circ$, and $\vartheta_3=45.1^\circ$.
The system operates at a carrier frequency of 2.4 GHz, corresponding to a wavelength of $\lambda = 0.125$ m.
Moreover, the transmit power of all users is set to the same value, i.e., $\bar{p}=p_1=...=p_K=0$ dB, and the SNR is defined as $ \rho=10 \text{log}10(\bar{p}/\sigma^2)$ with $\sigma^2$ denoting the noise power.
Unless otherwise stated, we set $\theta_{\text{max}}=\frac{\pi}{6}$, $p=4$, $T_{E}=60$, $M=6$, and $N=16$.

In the following, we present simulation results by averaging over 200 independent channel realizations. 
The performance is evaluated by the normalized mean square error (NMSE), defined as $\text{NMSE}= \mathbb{E} \left[ \frac{ \sum_{k=1}^{K}\left\| \boldsymbol{\eta}_k-\hat{\boldsymbol{\eta}}_k\right\|^2}{\sum_{k=1}^{K}\left\| \boldsymbol{\eta}_k\right\|^2}  \right]$,
where $\mathbf{\boldsymbol{\eta}}_k=
\beta_k \mathbf{a}_k(\vartheta_k,\phi_k)$ is the CSI determined only by environmental factors.
To validate the estimation performance advantages of our proposed RA-enabled system, we consider the following benchmark schemes for comparison:
1) \textbf{Random orientation design:} The deflection angles of each RA, i.e., $\left\lbrace \theta_{n}^\mathrm{z} \right\rbrace $ and $\left\lbrace \theta_{n}^{\mathrm{a}} \right\rbrace $ are randomly generated following a uniform distribution within $[0,\theta_{\text{max}}]$ and $[0,2\pi]$, respectively.
2) \textbf{Without orientation adjustment:} The orientations of all RAs are fixed as their reference orientations, i.e., $\mathbf{\Theta}=\mathbf{0}_{2\times N}$. 3) \textbf{Isotropic antenna system:} The directional gain is set to $G_0=1$ with $p=0$ in \eqref{radiation coefficient}. 
 \begin{figure}[!t]
 	\vspace{-0.3cm}
	\centering		
	\includegraphics[width=3.0in]{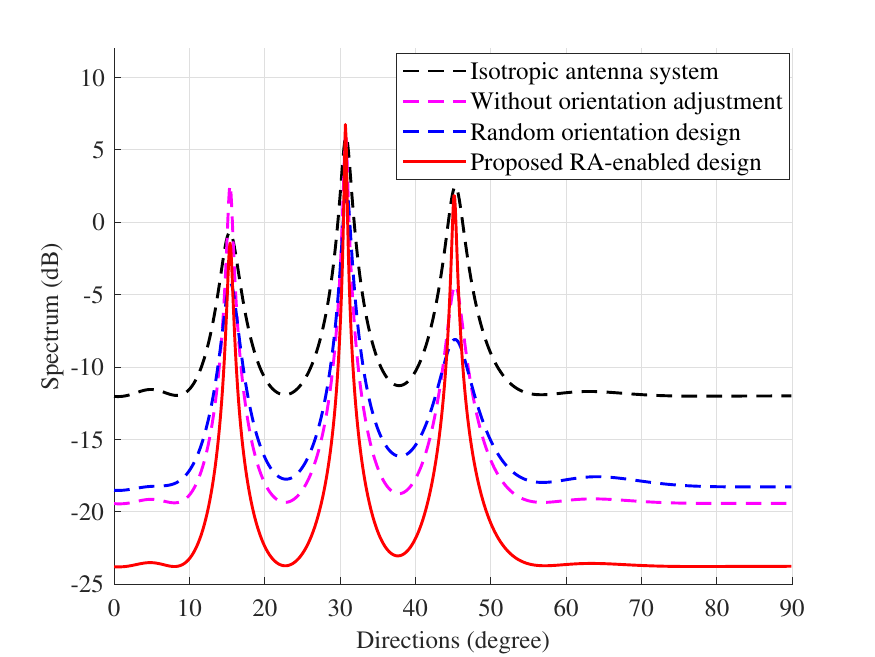}
    \vspace{-0.3cm}
	\caption{Angular spectrum versus user directions. 
		}
	\vspace{-0.4cm}
	\label{beampattern}
\end{figure}

In Fig. \ref{beampattern}, we show the resultant angular spectrum from four estimation schemes with $\text{SNR}=5$ dB. 
It is observed that compared to the benchmark schemes, the proposed RA-enabled scheme achieves narrower angular spectrum peaks directed towards user directions while maintaining significantly lower sidelobe levels.
This implies that the proposed RA-enabled scheme enhances angle resolution and estimation accuracy, which is attributed to the additional spatial DoFs and directional gains provided by RAs for scanning environments.

 \begin{figure}[!t]
	\centering		
	\includegraphics[width=3.0in]{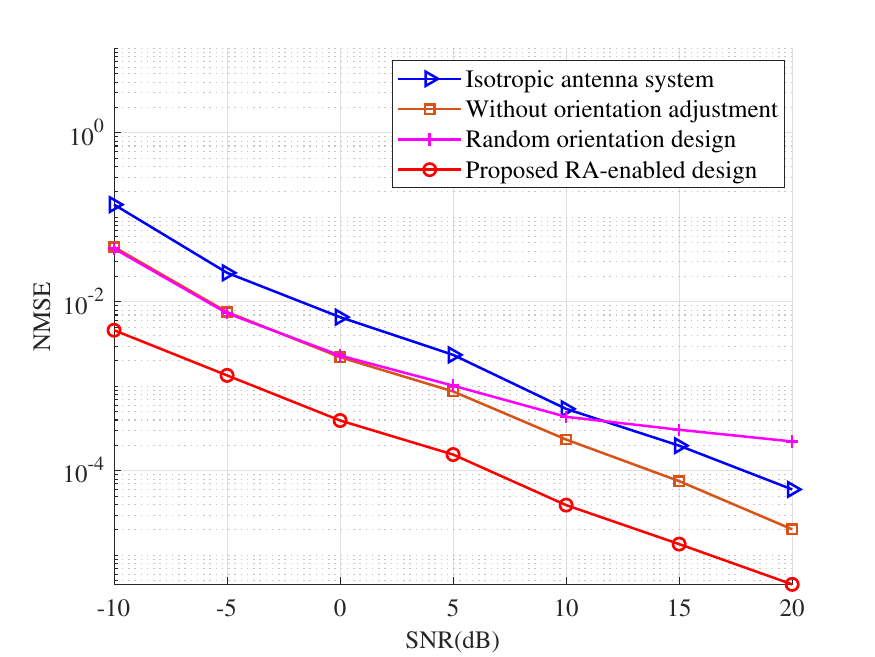}
	\vspace{-0.3cm}
	\caption{NMSE performance of the considered schemes versus SNR.}
	\label{MAE_SNR}
	\vspace{-0.1cm}
\end{figure}
In Fig. \ref{MAE_SNR}, we examine the channel estimation performance in terms of NMSE.
As the SNR increases, it can be observed that the NMSE of all schemes decreases. 
This is expected because the higher transmit power results in a stronger pilot signal received at BS, which is more favorable for channel estimation. On the other hand, across the entire SNR range, the proposed RA-enabled channel estimation method with preset orientations achieves lower NMSE compared to other estimation benchmarks.
This is due to the following two reasons. 
First, additional spatial DoFs introduced by RAs are leveraged to enrich the environmental observations from different directions, such that it can resolve more channel information.
Second, the RAs with meticulously designed orientations can enhance the system's directional capabilities, enabling more accurate multi-path resolution.

 \begin{figure}[!t]
	\centering
	\vspace{-0.3cm}		
	\includegraphics[width=3.0in]{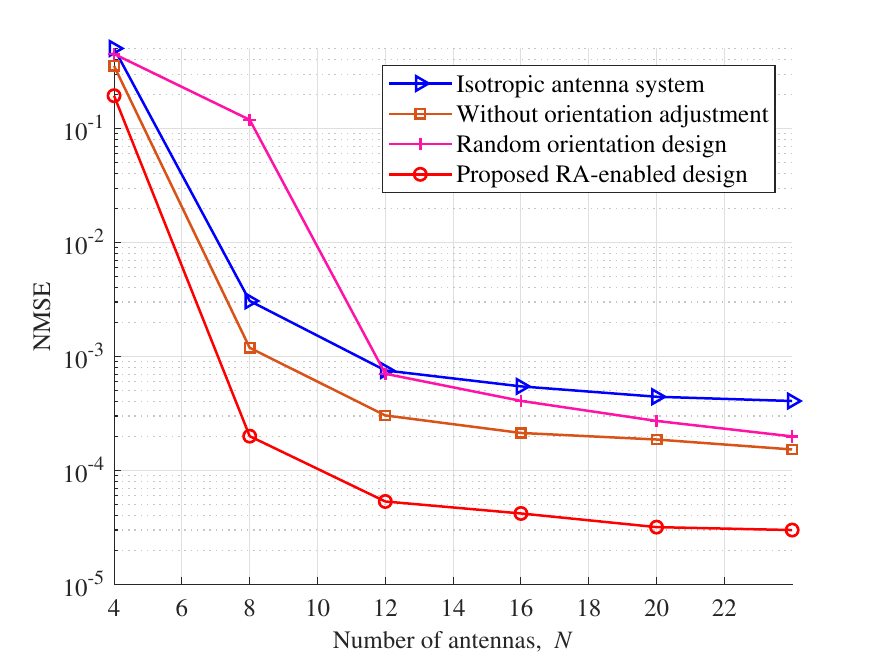}
	\vspace{-0.3cm}
	\caption{NMSE performance of the considered schemes versus the number of antennas at BS.}
	\label{MAE_RAs}
	\vspace{-0.5cm}
\end{figure}
In Fig. \ref{MAE_RAs}, we present the NMSE of different schemes versus the number of antennas. 
It is observed that, with an increase in the number of antennas, the NMSE for the proposed RA-based and other schemes decreases, which is expected due to the improved spatial resolution and array gain. 
Notably, the NMSE of the RA-based scheme remains lower than that of the other schemes. 
This performance gain stems from the enhanced angular discrimination and directional gain provided by the RAs, which effectively improve the received signal power and suppress interference from undesired directions.
In contrast, the schemes with fixed or random orientations suffer from misaligned radiation patterns, leading to suboptimal channel observability.
 
\vspace{-0.4cm}
\section{Conclusion}
In this letter, we considered the problem of channel estimation for the RA-enabled wireless communication system. 
To address this problem, we investigated an efficient estimation framework by exploiting the increased DoFs provided by RAs.
Specifically, the proposed scheme consisted of two main procedures: CSI estimation and RA orientation adjustment, which are conducted alternately during each channel training period.  
Simulation results have demonstrated that the proposed estimation scheme significantly improves the channel estimation accuracy.
%\vspace{-0.2cm}

%\bibliographystyle{IEEEtran}
%% argument is your BibTeX string definitions and bibliography database(s)
%\bibliography{RA-DOA}
\bibliographystyle{IEEEtran}
\bibliography{RA_CE}

\end{document}